\documentclass[a4paper]{article}
\usepackage{amsmath}
\begin{document}

\title{\textbf{On Two Aspects of the Painlev\'{e} Analysis}}
\author{\textsc{Sergei Sakovich}\\[12pt]
\textit{Institute of Physics, National Academy of Sciences,}\\
\textit{220072 Minsk, Belarus}\\[6pt]
\texttt{E-mail: saks@tut.by}}
\date{}
\maketitle

\begin{abstract}
We use the Calogero equation to illustrate the following two aspects of the
Painlev\'{e} analysis of nonlinear PDEs. First, if a nonlinear equation passes
the Painlev\'{e} test for integrability, the singular expansions of its
solutions around characteristic hypersurfaces can be neither single-valued
functions of independent variables nor single-valued functionals of data.
Second, if the truncation of singular expansions of solutions is consistent,
the truncation not necessarily leads to the simplest, or elementary,
auto-B\"{a}cklund transformation related to the Lax pair.
\end{abstract}

\section{Introduction}

The Painlev\'{e} analysis is a simple and reliable tool for testing the
integrability of nonlinear ODEs and PDEs \cite{RGB,Tab}. Concerning PDEs, there
is strong empirical evidence---though no proven theorem as yet---that every
nonlinear equation possessing the Painlev\'{e} property in formulation for PDEs
inevitably belongs either to the class of $C$-integrable equations (exactly
linearizable equations, including Darboux integrable ones) or to the class of
$S$-integrable equations (Lax integrable equations, including Liouville
integrable bi-Hamiltonian ones). The Painlev\'{e} analysis of nonlinear PDEs is
usually performed along the so-called Weiss--Kruskal algorithm, which combines
the Weiss' singular expansions of solutions around movable hypersurfaces
\cite{WTC} and the Kruskal's simplifying representation for singularity
manifolds \cite{JKM}, and which follows step by step the Ablowitz--Ramani--Segur
algorithm for ODEs \cite{ARS}.

The very first step of the Weiss--Kruskal algorithm, however, has no
counterpart in the Ablowitz--Ramani--Segur algorithm: starting the Painlev\'{e}
analysis of a nonlinear PDE, one must determine which of analytic hypersurfaces
are characteristic for the tested equation, in order to perform the whole
subsequent analysis of solutions around non-characteristic hypersurfaces only.
Ward \cite{War} was first who stated and substantiated that the Painlev\'{e}
property for PDEs must not fix any structure of solutions at characteristic
hypersurfaces. Afterward, the essence of Ward's statement was mentioned as
``a fact tacitly assumed by all Painlev\'{e} practitioners'' \cite{RGB}.
Lately, however, Weiss \cite{W1,W2} declared that his result ``runs counter to
the observation of Ward'' and that ``expansions about characteristic manifolds
are required to be single-valued'' as functionals of data.

In the present paper, in Section \ref{s2}, we show that the Ward's definition
of the Painlev\'{e} property for PDEs still remains well-founded and that the
objections of Weiss are caused by some terminological confusion. We do this via
the singularity analysis of the Calogero equation \cite{C75,CD}:
\begin{equation}
u_{xxxy}-2u_{y}u_{xx}-4u_{x}u_{xy}+u_{xt}=0 . \label{1}
\end{equation}

This nonlinear PDE (\ref{1}) is useful to illustrate one more aspect of the
Painlev\'{e} analysis. In Section \ref{s3}, we find two different B\"{a}cklund
transformations of the Calogero equation (\ref{1}) into itself: one follows
from the truncated singular expansion for $u$, the other one follows from the
Lax pair of (\ref{1}), and the former turns out to be a special case of the
square of the latter. Consequently, the Painlev\'{e} analysis does not lead to
the simplest, or elementary, auto-B\"{a}cklund transformation of (\ref{1}), a
phenomenon similar to what was observed in \cite{S94}.

Section \ref{s4} contains concluding remarks.

\section{Breaking Solitons and the Painlev\'{e} Property \label{s2}}

Let us take the fourth-order three-dimensional nonlinear PDE (\ref{1}) and
assume for a while that we know nothing about its integrability and solutions.
Does this equation (\ref{1}) pass the Painlev\'{e} test for integrability? The
answer will be ``yes'', if we adopt the Ward's definition \cite{War} of the
Painlev\'{e} property for PDEs. But the answer will be ``no'', if we change the
definition as proposed by Weiss \cite{W1,W2}.

It is an easy task to perform the Painlev\'{e} analysis of (\ref{1}) along the
Weiss--Kruskal algorithm. A hypersurface $\phi(x,y,t)=0$ is non-characteristic
for the PDE (\ref{1}) if $\phi_{x}^{3}\phi_{y}\neq0$ (see, e.g., \cite{Olv} for
the definition and meaning of non-characteristic hypersurfaces). The Kruskal's
ansatz $\phi=x+\psi(y,t)$ with $\psi_{y}\neq0$ both simplifies calculations and
excludes characteristic hypersurfaces from consideration. The assumption that
the dominant behavior of solutions is algebraic around $\phi=0$, that is
$u=u_{0}(y,t)\phi^{p}+\cdots$, leads to the only branch to be tested: $p=-1$
with $u_{0}=-2$. (Branches with $p=0,1,2,3$, also admitted by (\ref{1}), need
no analysis: they are governed by the Cauchy--Kovalevskaya theorem \cite{Olv}
because the Kovalevskaya form of the PDE (\ref{1}) is analytic everywhere.)
Then we substitute $u=-2\phi^{-1}+\cdots+u_{r}(y,t)\phi^{r-1}+\cdots$ into
(\ref{1}), find that $u_{r}$ is not determined if $r=-1,1,4,6$ ($r=-1$
corresponds to the arbitrariness of $\psi$), and conclude that the tested
branch is generic. Finally, we substitute $u=\sum_{i=0}^{\infty}u_{i}(y,t)
\phi^{i-1}$ into (\ref{1}), find recursion relations for $u_{i}$, and
check compatibility conditions at the resonances, where the arbitrary functions
$u_{1}$, $u_{4}$ and $u_{6}$ appear. All the compatibility conditions turn out
to be identities. Consequently, the PDE (\ref{1}) has passed the Weiss--Kruskal
algorithm well. Since this algorithm is sensitive to algebraic and non-dominant
logarithmic singularities only, we can only conjecture that the tested equation
possesses the Painlev\'{e} property in the sense that all solutions of (\ref{1})
are single-valued around all non-characteristic hypersurfaces. And we should
expect (\ref{1}) to be integrable.

The PDE (\ref{1}) is integrable indeed \cite{C75,CD}. It arises as the
compatibility condition for the over-determined, linear in $\Phi$, system
\begin{gather}
\Phi_{xx}+(\alpha-u_{x})\Phi=0 , \label{2}\\
\Phi_{t}+4\Phi_{xxy}-2u_{y}\Phi_{x}-4u_{x}\Phi_{y}-3u_{xy}\Phi=0 , \label{3}
\end{gather}
where the spectral parameter $\alpha$ is any function $\alpha(y,t)$ satisfying
the equation
\begin{equation}
\alpha_{t}=4\alpha\alpha_{y} . \label{4}
\end{equation}
All solutions of (\ref{4}), except $\alpha=\text{constant}$, are multi-valued
functions: for any non-constant initial value $\alpha(y,0)$, the nonlinear
wave $\alpha=\alpha(y,t)$ inevitably breaks (overturns, overlaps) at some
finite $t$ \cite{Whi}. Therefore solutions of (\ref{1}), obtainable by the
inverse scattering transform with any non-constant $\alpha$, are multi-valued
functions as well. For example, the one-soliton solution of (\ref{1}),
\begin{equation}
u=-2\lambda\tanh(\lambda x+\mu)+\beta , \label{5}
\end{equation}
where $\lambda(y,t)$, $\mu(y,t)$ and $\beta(y,t)$ are any functions satisfying
the equations $\lambda_{t}+4\lambda^{2}\lambda_{y}=0$ ($\lambda^{2}=-\alpha$)
and $\mu_{t}+4\lambda^{2}\mu_{y}=2\lambda\beta_{y}$, becomes a multi-valued
function when $\alpha$ breaks \cite{Bog}. The $N$-soliton solution of
(\ref{1}), determined by $N$ solutions $\alpha_{1},\ldots,\alpha_{N}$ of
(\ref{4}), breaks whenever any of $\alpha_{1},\ldots,\alpha_{N}$ breaks
\cite{Bog}.

At first sight, such a complicated branching of solutions of (\ref{1}) seems
to be incompatible with the Painlev\'{e} property. Nevertheless, there is no
contradiction between the fact that solutions of (\ref{1}) are multi-valued
functions and the fact that solutions of (\ref{1}) are single-valued around
all non-characteristic hypersurfaces: solutions can branch and do branch at
characteristic hypersurfaces only. Indeed, it was noticed and stressed in
\cite{Bog} that solutions of (\ref{1}) break (i.e. $u_{y}\rightarrow
\infty$ at finite values of $u$) for all values of $x$ simultaneously, the
fact meaning that the corresponding singularity manifolds $\phi=0$ are
characteristic for (\ref{1}), $\phi_{x}=0$. Consequently, the breaking
solitons do not break the Painlev\'{e} property in the Ward's formulation
\cite{War} because they never break at non-characteristic hypersurfaces.

Let us proceed to the Weiss' objections \cite{W1,W2} against the Ward's
formulation of the Painlev\'{e} property for PDEs. The Weiss' counter-example
is ``the expansion about the \textit{characteristic manifold}''
$u=u_{0}(t)+\sum_{i=3}^{\infty}u_{i}(t)\phi^{i}$ with $\phi=x+\psi(t)$ for the
equation $u_{xxx}=\frac{3}{2}u_{x}^{-1}u_{xx}^{2}+ku_{x}-u_{t}$, where
$k=\text{constant}$. This expansion, however, does not represent solutions
around characteristic hypersurfaces: characteristic hypersurfaces for this
equation are determined by the condition $\phi_{x}=0$, not by $\phi_{x}=1$.
Actually, this Taylor expansion represents solutions around any
non-characteristic hypersurface $\phi=0$ in the important case when the
Cauchy--Kovalevskaya theorem \cite{Olv} does not work: in this case we have
$u_{x}=0$ at $\phi=0$, whereas the Kovalevskaya form of the equation is
singular at $u_{x}=0$. We can only agree with Weiss that the consideration of
such special Taylor expansions is an essential part of the Painlev\'{e}
analysis, but the Weiss' words ``the expansion about the
\textit{characteristic manifold}'' turn out to be too misleading because no
actual expansions around characteristic hypersurfaces can be found in the
papers \cite{W1,W2}.

Now let us return to the Calogero equation (\ref{1}) and see what really
happens at the characteristic hypersurfaces $\phi(x,y,t)=0$ with
$\phi_{x}\phi_{y}=0$. When, for example, $\phi_{x}=0$ and $\phi_{y}\neq0$, we
take $\phi=y+\psi(t)$ and $u=u_{0}(x,t)\phi^{p}+\cdots$, $p=\text{constant}$,
and find from (\ref{1}) that any value of $p$ is admissible. Therefore the
expansions will not be single-valued functionals of $\phi$ for non-integer $p$.
For example, if $p=-\frac{1}{2}$, we get the expansion
\begin{equation}
u=\sum_{i=0}^{\infty}u_{i}(x,t)\phi^{(i-1)/2} \label{6}
\end{equation}
with the coefficients $u_{i}$ determined by the recursion relations
\begin{gather}
\sum_{i=0}^{n}(i-1)[u_{i}(u_{n-i})_{xx}+2(u_{i})_{x}(u_{n-i})_{x}] \notag \\
-\frac{1}{2}(n-2)[(u_{n-1})_{xxx}+\psi_{t}(u_{n-1})_{x}]-(u_{n-3})_{xt}=0 ,
\label{7}
\end{gather}
where $n=0,1,2,\ldots$, and $u_{i}=0$ at $i<0$. The structure of (\ref{7})
differs from the habitual structure of recursion relations for
non-characteristic hypersurfaces very considerably. There are no resonances in
(\ref{7}), but the expansion (\ref{6}) contains infinitely many arbitrary
functions of $t$ in addition to $\psi(t)$: they arise pair by pair as
``constants'' of integration of (\ref{7}) because (\ref{7}) is a second-order
ODE in $u_{n}$ for every $n$. Namely, $u_{0}=(\sigma_{0}x+\tau_{0})^{1/3}$,
$u_{1}=\frac{5}{36}\sigma_{0}(\sigma_{0}x+\tau_{0})^{-1}+\sigma_{1}
(\sigma_{0}x+\tau_{0})^{1/3}+\tau_{1}+\frac{1}{4}\psi_{t}x$,
$u_{2}=\sigma_{2}(\sigma_{0}x+\tau_{0})^{1/3}+\tau_{2}(\sigma_{0}x+
\tau_{0})^{2/3}$, etc., where $\sigma_{i}(t)$ and $\tau_{i}(t)$ are arbitrary
functions, $i=0,1,2,\ldots$. We see that the expansion (\ref{6}) is
multi-valued both as a function of $x,y,t$ (via coefficients $u_{i}$ and
non-integer degrees of $\phi$) and as a functional of $\phi$. Consequently, if
we accept the Weiss' formulation \cite{W1,W2} of the Painlev\'{e} property, the
integrable Calogero equation (\ref{1}) will not pass the Painlev\'{e} test for
integrability. Evidently, the Weiss' formulation asks too much of the tested
equation.

\section{Two auto-B\"{a}cklund transformations \label{s3}}

Let us try to find a B\"{a}cklund transformation of the Calogero equation
(\ref{1}) into itself. Two different methods will lead us to two different
transformations. Then we will find a relation between the two results.

First we employ the method of truncated singular expansions of Weiss
\cite{Wei} and the new expansion function $\chi=(\phi^{-1}\phi_{x}-
\frac{1}{2}\phi_{x}^{-1}\phi_{xx})^{-1}$ of Conte \cite{Con} which simplifies
computations very considerably (note also that the Kruskal's ansatz is not used
for $\phi$ in what follows). We substitute $u=g(x,y,t)\chi^{-1}+f(x,y,t)$ into
the Calogero equation (\ref{1}) and find that $g=-2$ and that $\phi$ and $f$
must satisfy the following system of four equations:
\begin{gather}
d-2c(s+2f_{x})+2f_{y}=0 , \label{8} \\
d_{x}-\frac{1}{2}c(s_{x}+2f_{xx})-2c_{x}(s+2f_{x})+2f_{xy}=0 , \label{9} \\
d_{xx}+ds-c_{x}(s_{x}+2f_{xx})-2(c_{xx}+cs)(s+2f_{x})-s_{xy}+2sf_{y}=0 ,
\label{10} \\[6pt]
s_{xxy}+f_{xxxy}+(c_{xx}+cs)(s_{x}+2f_{xx})-2f_{y}(s_{x}+f_{xx}) \notag \\
-4(s+f_{x})(s_{y}+f_{xy})+2ss_{y}+s_{t}+f_{xt}=0 , \label{11}
\end{gather}
where $s=\phi_{x}^{-1}\phi_{xxx}-\frac{3}{2}\phi_{x}^{-2}\phi_{xx}^{2}$,
$c=-\phi_{x}^{-1}\phi_{y}$, and $d=-\phi_{x}^{-1}\phi_{t}$. Substituting
(\ref{8}) into (\ref{9}), we get $s_{x}+2f_{xx}=0$ which leads to
$s+2f_{x}=2\alpha$, where the function $\alpha(y,t)$ appears as a ``constant''
of integration. Then (\ref{8}) changes into $d-4\alpha c+2f_{y}=0$, (\ref{10})
is satisfied identically, and $\alpha_{t}=4\alpha\alpha_{y}$ follows from
(\ref{11}) (that is why we use the same letter $\alpha$ as for the spectral
parameter). Consequently, the system (\ref{8})--(\ref{11}) is equivalent to the
system of two equations
\begin{equation}
\label{12}
\begin{gathered}
\phi_{xxx}-\frac{3}{2}\phi_{x}^{-1}\phi_{xx}^{2}+2\phi_{x}f_{x}-2\alpha
\phi_{x}=0 , \\
\phi_{t}-2\phi_{x}f_{y}-4\alpha\phi_{y}=0 ,
\end{gathered}
\end{equation}
where $\alpha(y,t)$ is any solution of (\ref{4}). The truncated expansion
\begin{equation}
u=\phi_{x}^{-1}\phi_{xx}-2\phi^{-1}\phi_{x}+f \label{13}
\end{equation}
is a Miura transformation of the system (\ref{12}) into the equation
(\ref{1}). One more Miura transformation of (\ref{12}) into (\ref{1}), namely,
\begin{equation}
v=\phi_{x}^{-1}\phi_{xx}+f , \label{14}
\end{equation}
where $v$ satisfies (\ref{1}), follows from (\ref{13}) automatically
\cite{Wei,Con}. This chain of two Miura transformations (\ref{13}) and
(\ref{14}) generates an auto-B\"{a}cklund transformation for (\ref{1}). Indeed,
eliminating $\phi$ and $f$ from (\ref{13}) and (\ref{14}) by means of
(\ref{12}) and differentiations, we get the following system:
\begin{gather}
w_{xx}-\frac{1}{2}w^{-1}w_{x}^{2}-wz_{x}+\frac{1}{8}w^{3}+2\alpha w=0 ,
\label{15} \\
z_{xy}-w^{-1}(w_{x}z_{y}+4\alpha w_{y}-w_{t})-\frac{1}{2}ww_{y}-4\alpha_{y}=0 ,
\label{16}
\end{gather}
where $w=u-v$, $z=u+v$, and $\alpha(y,t)$ is any solution of (\ref{4}). Direct
but tedious computations prove that the system (\ref{15})--(\ref{16}) is
compatible in $v$ if $u$ satisfies (\ref{1}), and that the system is
compatible in $u$ if $v$ satisfies (\ref{1}) (i.e. one gets (\ref{1}) for $u$
when eliminates $v$ from (\ref{15})--(\ref{16}) by differentiations, and vice
versa). Therefore, according to the definition \cite{AS}, the system
(\ref{15})--(\ref{16}) is a B\"{a}cklund transformation of the PDE (\ref{1})
into itself.

It looks strange, however, that the $x$-part (\ref{15}) of the obtained
B\"{a}cklund transformation is a second-order ODE, whereas the equation
(\ref{2}) of the associated linear problem for the PDE (\ref{1}) is the same
as for the potential KdV equation $u_{t}=u_{xxx}-3u_{x}^{2}$. Let us apply the
method of Chen \cite{Che} to the linear problem (\ref{2})--(\ref{3}) and find
that the PDE (\ref{1}) does admit one more auto-B\"{a}cklund transformation
with the same first-order $x$-part as for the potential KdV equation. We
rewrite (\ref{2}) as $u_{x}=(\Phi_{x}/\Phi)_{x}+(\Phi_{x}/\Phi)^{2}+\alpha$,
introduce the new variable $\omega$ such that $\omega_{x}=(\Phi_{x}/\Phi)^{2}+
\alpha$, and get in this way $u=\omega\pm\varepsilon$, where $\varepsilon
=(\omega_{x}-\alpha)^{1/2}$. Then (\ref{3}) gives us the following
fourth-order PDE for $\omega$:
\begin{equation}
\varepsilon_{xxy}-2(\omega_{y}\varepsilon)_{x}-4\alpha\varepsilon_{y}+
\varepsilon_{t}=0 . \label{17}
\end{equation}
It is very essential that (\ref{17}) is one and the same equation for both
choices of the sign in $u=\omega\pm\varepsilon$. Owing to this fact, we have
two Miura transformations of (\ref{17}) into (\ref{1}), namely,
\begin{align}
u &=\omega+(\omega_{x}-\alpha)^{1/2} , \label{18} \\
v &=\omega-(\omega_{x}-\alpha)^{1/2} , \label{19}
\end{align}
where $u$ and $v$ are solutions of (\ref{1}) if $\omega$ satisfies (\ref{17}).
Eliminating $\omega$ from (\ref{18}), (\ref{19}) and (\ref{17}), we get
\begin{gather}
z_{x}-\frac{1}{2}w^{2}-2\alpha=0 , \label{20} \\
w_{xxy}-w_{x}z_{y}-(w^{2}+4\alpha)w_{y}+w_{t}-2\alpha_{y}w=0 , \label{21}
\end{gather}
where $w=u-v$, $z=u+v$, and $\alpha(y,t)$ is any solution of (\ref{4}). One
can check that the system (\ref{20})--(\ref{21}) is compatible in $v$ if $u$
satisfies (\ref{1}), and vice versa. Therefore the equations (\ref{20}) and
(\ref{21}) constitute a B\"{a}cklund transformation of the PDE (\ref{1}) into
itself.

We have obtained two auto-B\"{a}cklund transformations for the
Calogero equation (\ref{1}): (\ref{15})--(\ref{16}) and (\ref{20})--(\ref{21}).
The two transformations are different both in their form, what is evident, and
in solutions $u$ they generate from a given solution $v$. For example, if $v$
is any function $\gamma(y,t)$, we find from (\ref{20})--(\ref{21}) that the
corresponding $u$ is the one-soliton solution (\ref{5}) with $\beta=\gamma$,
whereas the transformation (\ref{15})--(\ref{16}) leads for this $v$
either to the soliton (\ref{5}) with $\beta=\gamma-2\lambda$ or to the more
complicated solution
\begin{equation}
u=-8\lambda[\cosh(\lambda x+\mu)]^{2}\{\sinh[2(\lambda x+\mu)]+2(\lambda
x+\nu) \}^{-1}+\gamma , \label{22}
\end{equation}
where the functions $\lambda(y,t)$, $\mu(y,t)$ and $\nu(y,t)$ are any
solutions of the equations $\lambda_{t}+4\lambda^{2}\lambda_{y}=0$
($\lambda^{2}=-\alpha$), $\mu_{t}+4\lambda^{2}\mu_{y}=2\lambda\gamma_{y}$ and
$\nu_{t}+4\lambda^{2}\nu_{y}-8\lambda\lambda_{y}\nu=2\lambda\gamma
_{y}-8\lambda^{2}\mu_{y}$. (For completeness, we should also mention the
generated solutions $u$ with $u_{x}=0$: $u=\gamma-2\lambda$ for
(\ref{20})--(\ref{21}), and $u=\gamma$ and $u=\gamma-4\lambda$ for
(\ref{15})--(\ref{16}), $\lambda^{2}=-\alpha$.) Nevertheless, these two
different auto-B\"{a}cklund transformations are related to each other: the
transformation (\ref{15})--(\ref{16}) is nothing but a special case of the
square of the transformation (\ref{20})--(\ref{21}). More precisely, if
functions $a$, $b$ and $q$ are such that $u=a$ and $v=q$ satisfy the system
(\ref{20})--(\ref{21}) with some spectral parameter $\alpha$, and $u=q$ and
$v=b$ satisfy the system (\ref{20})--(\ref{21}) with the same $\alpha$, then
$u=a$ and $v=b$ satisfy the system (\ref{15})--(\ref{16}) with the same
spectral parameter $\alpha$. Indeed, eliminating $q$ from the relations
$a_{x}+q_{x}=\frac{1}{2}(a-q)^{2}+2\alpha_{1}$ and $q_{x}+b_{x}=\frac{1}{2}
(q-b)^{2}+2\alpha_{2}$, we get (\ref{15}) for $u=a$ and $v=b$ if and only if
$\alpha_{1}=\alpha_{2}=\alpha$, and (\ref{16}) follows from (\ref{21}) in the
same way. Therefore our words ``a special case of the square'' mean that the
transformation (\ref{15})--(\ref{16}) is composed of two transformations
(\ref{20})--(\ref{21}) with equal spectral parameters.

We have shown that the method of truncated singular expansions does not lead to
the simplest auto-B\"{a}cklund transformation of the Calogero equation
(\ref{1}), related to the equation's Lax pair, and one may only guess that
the transformation (\ref{20})--(\ref{21}) can be derived from the
transformation (\ref{15})--(\ref{16}).

\section{Conclusion\label{s4}}

In this paper, we used the Calogero equation to illustrate the following two
aspects of the Painlev\'{e} analysis of nonlinear PDEs.

First, if a nonlinear equation passes the Painlev\'{e} test for integrability,
the singular expansions of its solutions around characteristic hypersurfaces
can be neither single-valued functions of independent variables nor
single-valued functionals of data. Of course, if the Painlev\'{e} property is
considered as an abstract analytic property, one may give any definition of it.
However, if the Painlev\'{e} property is defined to be used as an indicator of
integrability of nonlinear equations, the adequacy of its definition becomes an
experimental result. By the singularity analysis of the Calogero equation, we
have shown that the Ward's definition of the Painlev\'{e} property for PDEs is
well founded.

Second, if the truncation of singular expansions of solutions is consistent,
the truncation not necessarily leads to the simplest, or elementary,
auto-B\"{a}cklund transformation related to the Lax pair. We have found two
different B\"{a}cklund transformations of the Calogero equation into itself:
one follows from the truncated singular expansion, the other one follows from
the Lax pair, and the former turns out to be a special case of the square of
the latter. In other words, the way from the truncated singular expansions to
B\"{a}cklund transformations and Lax pairs is not so straightforward as it is
sometimes stated in the literature.

\end{document}